\documentclass[a4paper,12pt,oneside]{article}
\usepackage[intlimits]{amsmath}
\usepackage{amsfonts,amssymb}
\usepackage{wrapfig}
\usepackage[dvips]{graphicx}
\usepackage{psfrag}
\usepackage{bm}
\usepackage{amsmath}

\def\tN{\mathop{\rm N}\nolimits}

\makeindex
\DeclareSymbolFont{letterg}{OT1}{cmr}{m}{sl}\DeclareMathSymbol{g}{\mathalpha}{letterg}{`g}

\begin{document}

{\Large
{
\begin{center}
\textbf{Consideration of arrayed $e$-beam microcolumn based systems
potentialities for wafer defects inspection}
\end{center}
}}

\begin{center}
\textit{V.V. Kazmiruk, T.N. Savitskaja}
\end{center}

{\footnotesize
{\begin{center}
\textit{Institute of Microelectronics Technology
and High Purity Materials, Russian Academy of~Sciences}
\end{center}
}}
\vskip4pt

\begin{center}
\textbf{Abstract}
\end{center}

{\small
{
The $e$-beam column which is intended on defects inspection is considered.
The defects which are to be
examined or potentially might be examined at inspection stage
are briefly considered.
Interrelations between the system parameters is ascertaining
and the ways of optimization and
the technical requirements to the system in whole are discussed.
As a result, we find the optimal combinations of the system parameters for the purpose.
}}

{\large
{
\begin{center}
\textbf{Introduction}
\end{center}
}}

The development and manufacture of solid state microstructures involves many
intricate processes for their creating. Along with the highly developed
technologies used to manufacture micro and nanostructures, there is need for
testing processes which would be adequate to those technologies by
resolution, sensitivity and throughput. This is the increasingly important
to shorten product development cycles and to increase the product yield.

Basically, test processes comprise inspection and review stages. The first
one is used to detect, identify and locate defects (or potential defects) on
wafer After the potential defects are located a review stage is conducted.
As usually the review process involves much more detailed examination of
individual defects. For instance, the size, shape, nature and cause of
defect can be determined.

The key question should be answered by inspection process is total amount of
<<killer>> defects on wafer and possibility to return the wafer for further
processing if is assumed some tolerable probability to have at output an
acceptable yield.

The aim of review stage is to determine the cause of the defect appearance
in order to improve a technology and exclude such defects in a future, or at
least decrease their amount. Usually this stage does not contemplate to
return the wafer into technology process. Often it is used in combination
with other techniques as for instance X-ray microanalysis, selective ion
etching performed by FIB and so on.

At the present time for defect inspection are using the systems based on
light optics. These are highly automated systems with inspection speed about
1 cm$^{2}$ in 10 sec, which means approximately one wafer in a hour.
However, their resolution is limited by 0,25~$\mu $m.

As electron beam can be focused into spot a few nanometers only, a solutions
seems to be obvious: just to replace the light optics on $e$-beam column,
taken from a SEM as for instance.

However, after more detail consideration of that idea one can see three main
obstacles on this way. The first consists in the principle of implemented
method for image formation, eg in pixel-by-pixel scanning. Lowering the
pixel size we quadratically decrease an inspection speed.

The second obstacle is that differently from the light, an $e$-beam is
interacting with a sample, which might be a cause of radiation damage. On
the other hand, some SEM methods are based on electron scattering
computation to extract useful information, so that phenomena can play
positive role either.

And the third, that whole image formation process, starting from the beam
generation and ending by signal detection, has a statistical character. This
also can put some limitations on the inspection system characteristics.

All that means that for creation of the system, and even more so its
optimization, it is necessary to describe the whole system behavior,
including above mentioned factors. Just after that is possible to synthesize
an optimal system.

Often, when estimating a SEM, are considering just one its parameter~--- the
resolution. Some time is added accelerating voltage, and then SEM is called
something like "Low voltage high resolution SEM" et cetera. And the best
SEM in resolution is assumed to be the best for any application. In this
work we have put at the first place the application.

In other words, the aim is to create the system which is focused on concrete
application~--- defects inspection, and with such combination of parameters
that to the maximum adapted for this particular task.

In the first part of the work are considering briefly defects which to be
examined or potentially might be examined at inspection stage.

In the second part  interrelations between the major system
parameters is ascertaining. The ways of optimization and technical
requirements to the both components and system at whole are discussed.
\vskip15pt

{\large
{\begin{center}
\textbf{I. The structures parameters}
\end{center}
}}
\vskip8pt

Solid state micro and nanostructures are described by 3 groups of
parameters:

\noindent
--- geometrical dimensions and configuration of the structure's elements;

\noindent
--- metallurgical parameters as elemental composition, distribution of
impurities and defects on wafer and so on;

\noindent
--- local electrical parameters as concentration of free charge carriers, their
mobility and life time, dielectrical constants and so on.

The process of production of semiconductors includes processing of a
circular silicon wafer typically 8" in diameter. The processing includes
repetition of series of steps: oxidation and deposition; lithography;
etching and doping (implanting and diffusing).

\begin{center}
\includegraphics[scale=1.1]{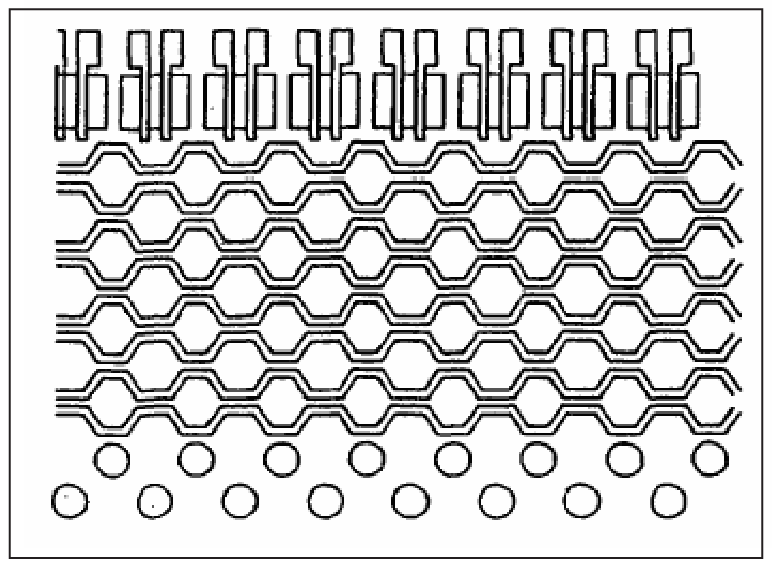}
\end{center}

\centerline{Fig.\:1 \textit{a}. Normally fabricated pattern}

\begin{center}
\includegraphics[scale=1.1]{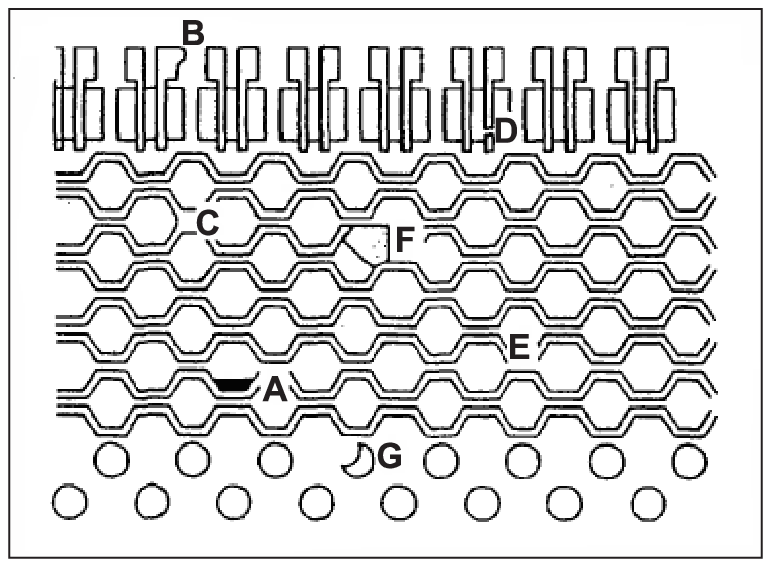}
\end{center}

$\phantom{a}$
\vspace*{-14mm}

\begin{center}
{Fig.\:1 \textit{b}. Pattern having a defect \textbf{A: }isolated defect\textbf{ B}:\par protrusion \textbf{C}: short \textbf{D}: omission \par \textbf{E}\textbf{:} disconnection
\textbf{F}: thin film residual \textbf{G}: bad aperture }
\end{center}

\noindent
 Depending on the maturity of the production process used, the
wafer might be inspected for particles/production defects, mask alignment
and critical dimension metrology between the processing steps. The frequency
of inspection can be as often as every wafer in the development phase of a
process, or on wafers from alternate production lots from mature processes.

Particle (production defect) detection detects either the presence of
contaminant particles introduced in the manufacturing process, or areas
where processing has been defective so as to produce unwanted features in
the structures of the device. An example of defects caused by manufacturing
process is given at fig.\:1. These are so-called topographical defects.

SEM-based inspection systems for this kind of defects have been proposed
using die-to-die comparison methods. Such systems are optimized to obtain
topographical information. Known techniques have small pixel size (0,1~$\mu
$m) and consequently very long inspection times, of the order of 10 to 80
hours for a complete wafer.

Two more defects caused by $e$-beam
lithography are shown on fig.\:2 and fig.\:3.

\begin{table}[htbp]
\begin{tabular}
{|p{180pt}|p{180pt}|}
\hline
\centerline{\includegraphics[scale=0.85]{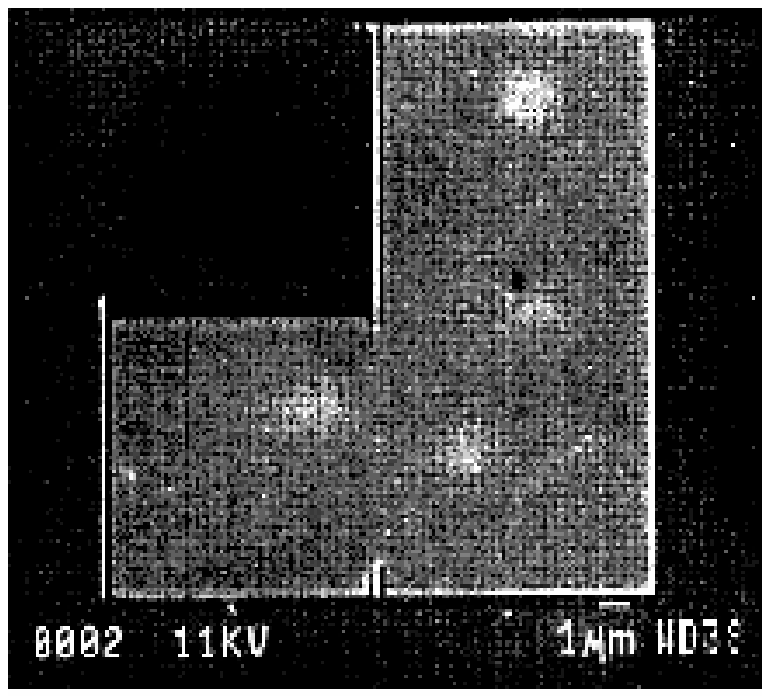}}&
\centerline{\includegraphics[scale=0.85]{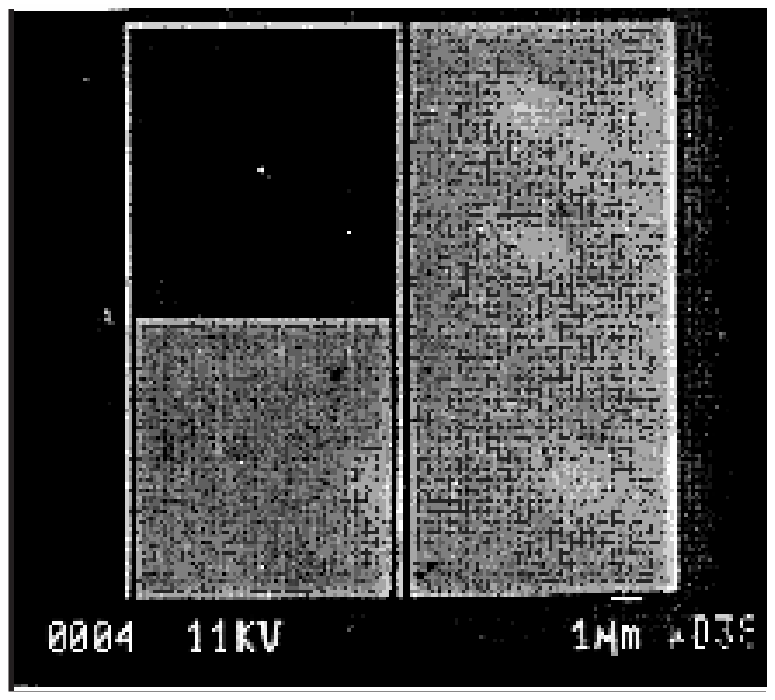}}\\
\hline
 \par Fig. 2\,\textit{a}. Transistor without proximity effect correction&
Fig. 2\,\textit{b}. After proximity effect correction implemented \\
\hline
\end{tabular}
\label{tab2}
\end{table}

At fig.\:2\,\textit{a} is shown top plan view of FET transistor which has been exposed
without including into consideration an existence of proximity effect. As a
result the gap between drain and source has been over exposed. After
developing process and metal deposition has been formed short circuit
between all the electrodes. At fig.\:2\,\textit{b} is shown the same kind of transistor
after problem was fixed.

At fig.\:3\,\textit{a}, \textit{b} is shown a test structure
to study $e$-beam lithography defects
caused by hysteresis of deflection system. After all the temporary delays
have been included into account and compensated by programming means, then
was possible to achieve a desired result (fig.\:3\,\textit{b}).

\begin{table}[htbp]
\begin{tabular}
{|p{180pt}|p{180pt}|}
\hline
\centerline{\includegraphics[scale=1.3]{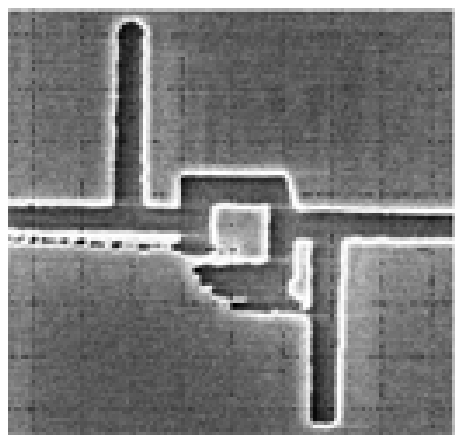}}&
\centerline{\includegraphics[scale=0.85]{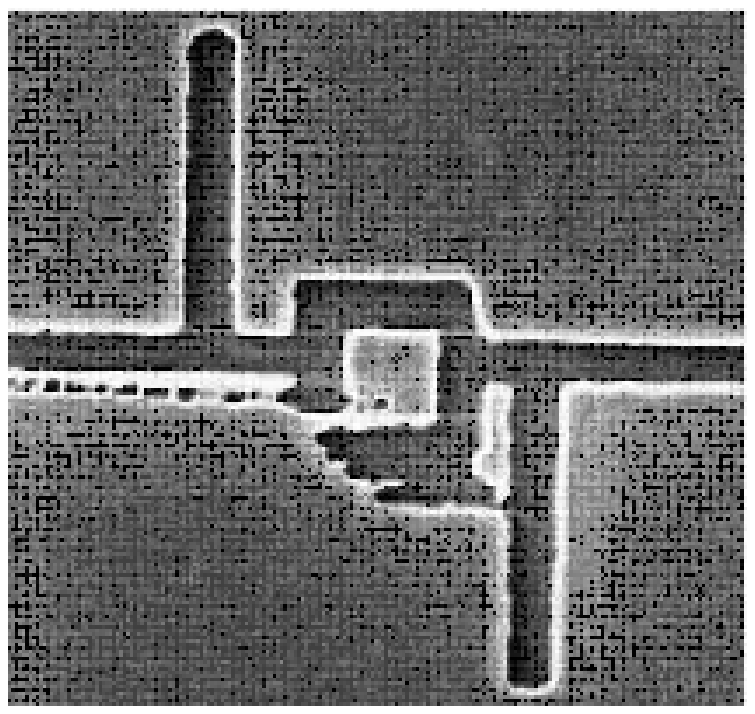}}\\
\hline
   \centerline{Fig.\:3 \textit{a}}&
 \centerline{Fig.\:3 \textit{b}} \\[-12pt]
\hline
\multicolumn{2}{|p{360pt}|}{* The explanations are in the text.}  \\
\hline
\end{tabular}
\label{tab3}
\end{table}
\par
\pagebreak

The defects (particles) can be subdivided into two groups: those which will
affect the operation of the completed structure or device, which are known
as "killer" \ defects, and those which have no harmful effect, known as
"nuisance" \ defects. As for instance, protrusions {\bf{B}} on fig.\:1\,\textit{b}
is a rather
nuisance defects, and all the rest are killer defects. The small particles
on transistor which are clearly seen on fig.\:2, are nuisance defects, and
both short circuit on fig.\:2\,\textit{a} and lithography
defects on fig.\:3\,\textit{a} are
"killer" \ defects.
On the other hand, the nuisance defects can account for 90\,{\%} of detected
defects; therefore, some form of review is required to ensure that wafers
which would otherwise produce acceptable yields are not rejected.

One more thing is that some of defects as thin film residuals are not
clearly seen at the SEM image. That can be both killer and nuisance defects.
Have been proposed the methods of such defects inspection based on voltage
contrast (US~patent {\#}7,132,301 B1) The idea is explained by fig.\:4, 5.

\begin{table}[htbp]
\begin{tabular}
{|p{180pt}|p{180pt}|}
\hline
\centerline{\includegraphics[scale=0.65]{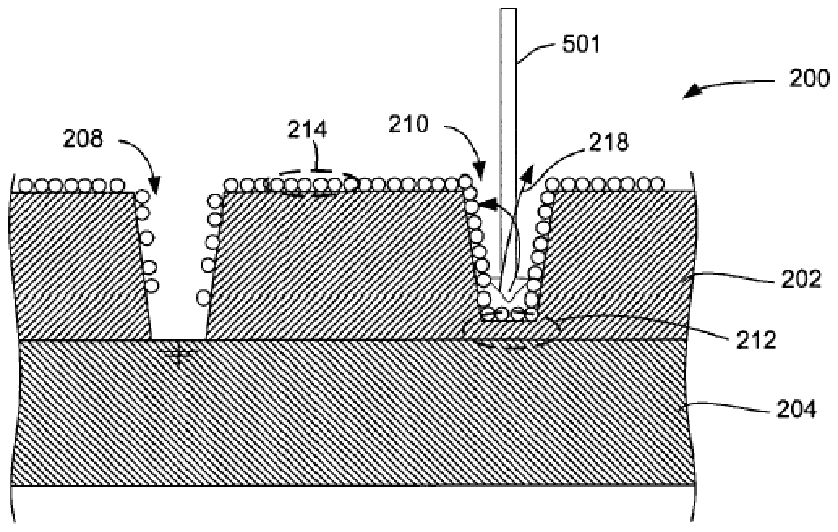}} \par &
\centerline{\includegraphics[scale=0.65]{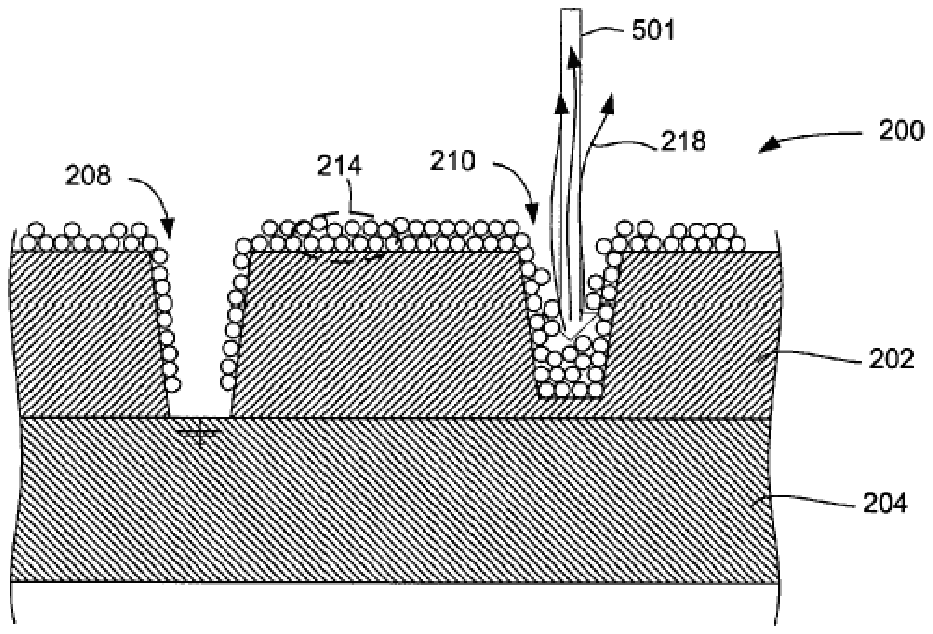}} \par  \\
\hline
Fig.\:4\,\textit{a}. Cross-sectional view of portion of wafer.&
Fig.\:4\,\textit{b}. The same area after been flooded. \\
\hline
\centerline{\includegraphics[scale=0.43]{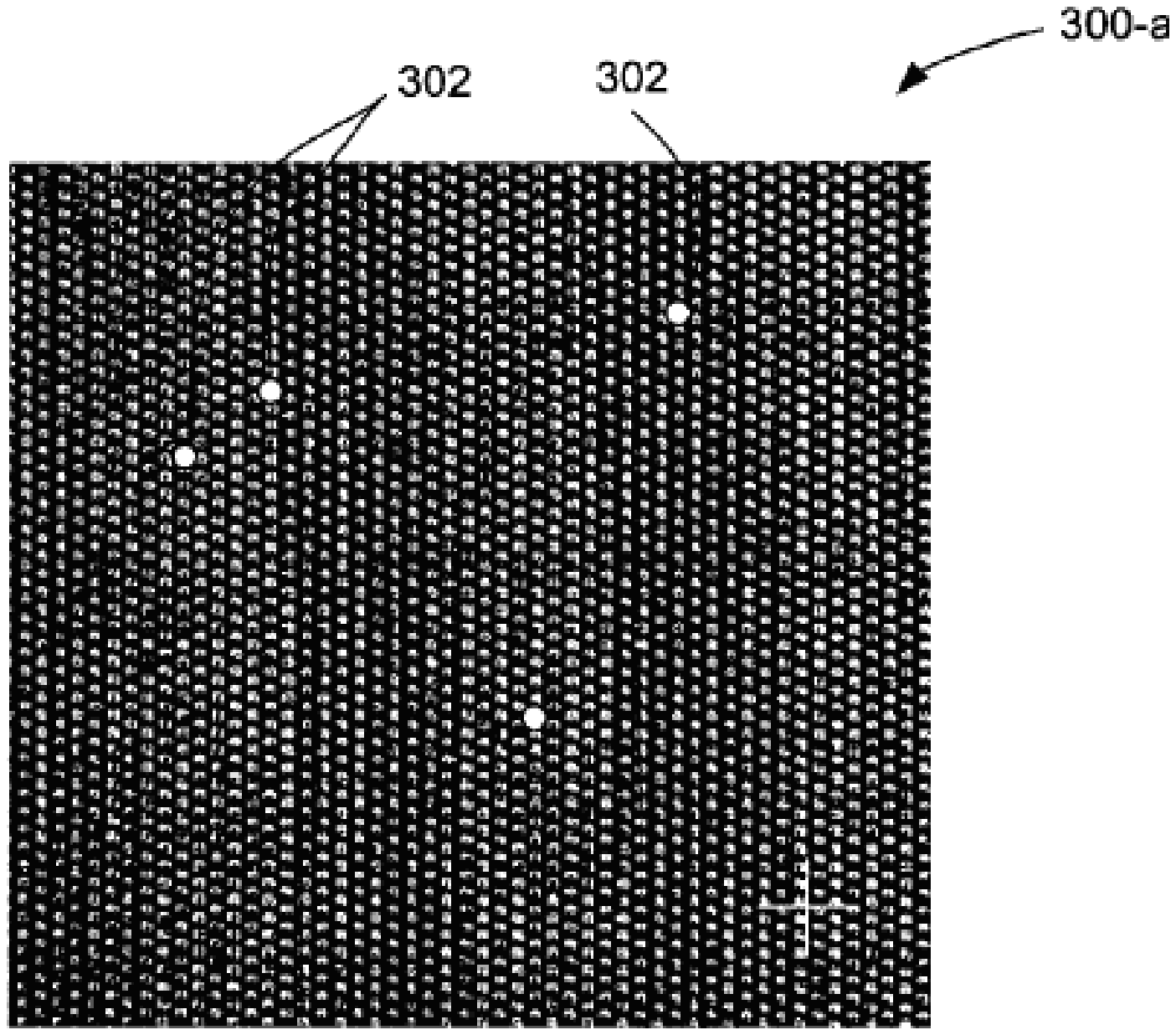}} \par &
\centerline{\includegraphics[scale=0.43]{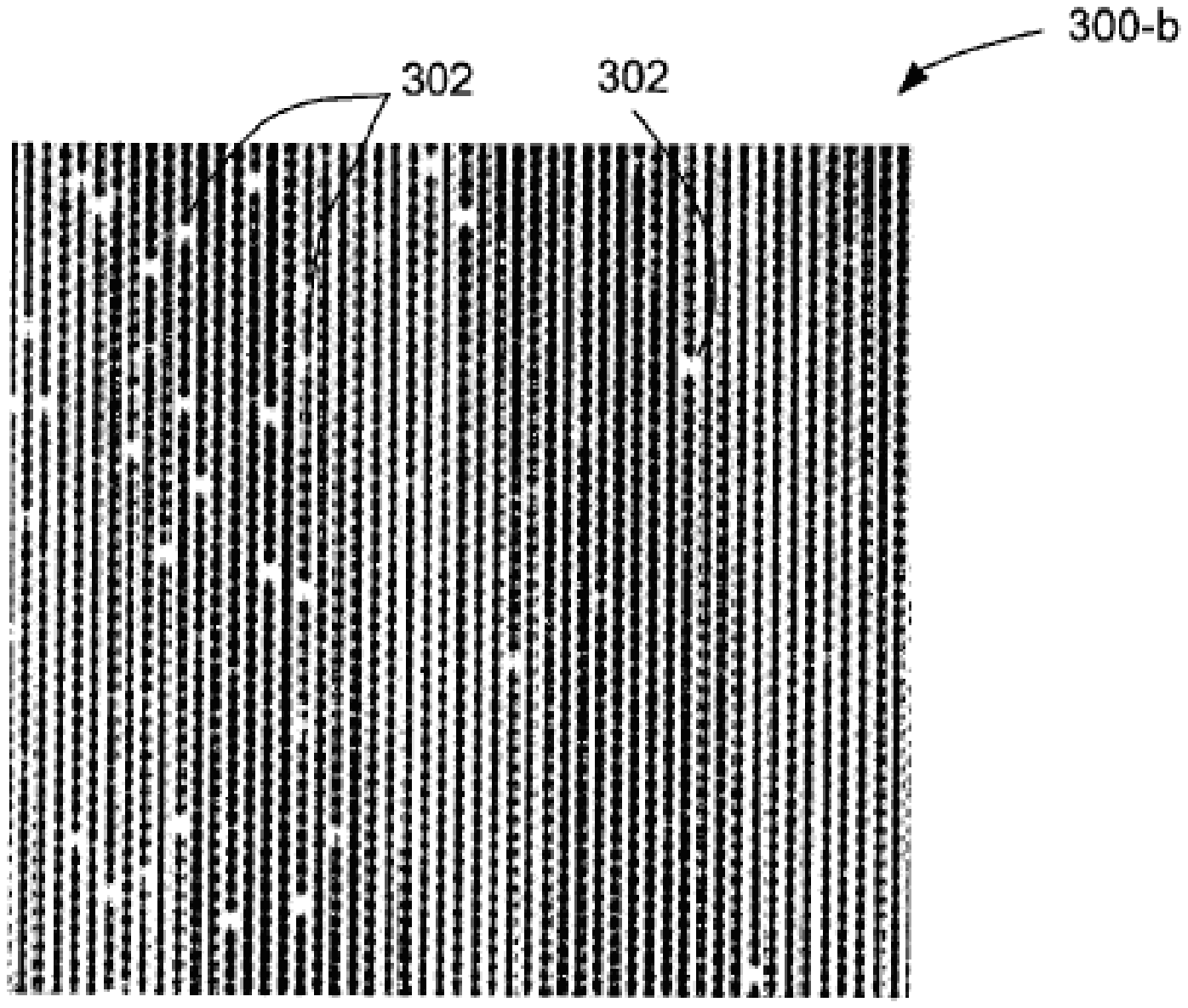}} \par  \\
\hline
Fig.\:5\,\textit{a}. Top view of the wafer without flooding.&
Fig.\:5\,\textit{b}. The same area after been flooded. \\
\hline
\multicolumn{2}{|p{360pt}|}{[1] Us patent {\#} 7,132,301 B1}  \\
\hline
\end{tabular}
\label{tab4}
\end{table}

If a focused $e$-beam is scanning over the wafer it is charging some existing
thin film residuals as it's shown diagrammatically on Fig. 2a. The value of
the charge depends on several parameters as film thickness, beam energy,
dose of irradiation an so on. In some places, where big enough negative
charge is occurred, on the SEM image are seen bright areas. Those areas are
potentially "weak" points where defects might exist or arise in a future.
The reason of such signal formation is that negative voltage on the surface
creates a local accelerating field between irradiated point and detector,
which directs the slow secondary electrons (SE) towards detector. This kind
of contrast is called quantitative voltage contrast. The voltage contrast
can be enhanced when using additionally one or more flooding beam. The
"flooding" means some process when large enough electron beam covers
homogeneously the whole wafer or inspected area. When the wafer has been
flooded (see fig.\:4\,\textit{b}), the negative charge is increased and many more weak
points become seen on the image (compare image fig.\:5\,\textit{a} and \textit{b}).

The main method of extracting information by pictures comparison is based on
the assumption that in a SEM just as in light microscope, the sample is
adequately represented by its image.

The commonly accepted definition for that kind of data is \textit{qualitative
information}, while a lot of statistical information can be achieved by image
analyzing. For example, these are number of particles and defects, their
distribution by size and that sort of things.

Another kind of information which is obtained from SEM signal directly or by
applying some algorithm commonly is defined as \textit{quantitative information}.

Leaving apart the discussion about details, we just note that qualitative
information is related to image and quantitative information is related to
the signal from a SEM detector.

One of the earliest and the most developed methods is critical dimensions
(CD) measurement or, according to another authors, line width (LW)
measurement.

One more example of defects is shown on fig.\:6, 7 and 8. These are structures
for X-ray focusing.
In ideal case the depth of all the trenches has to be the same. This
determines a X-ray quality. However, after ion etching one can clearly see
that trench depth is decreasing with decreasing its width.

\begin{figure}[htbp]
\centerline{\includegraphics[scale=0.9]{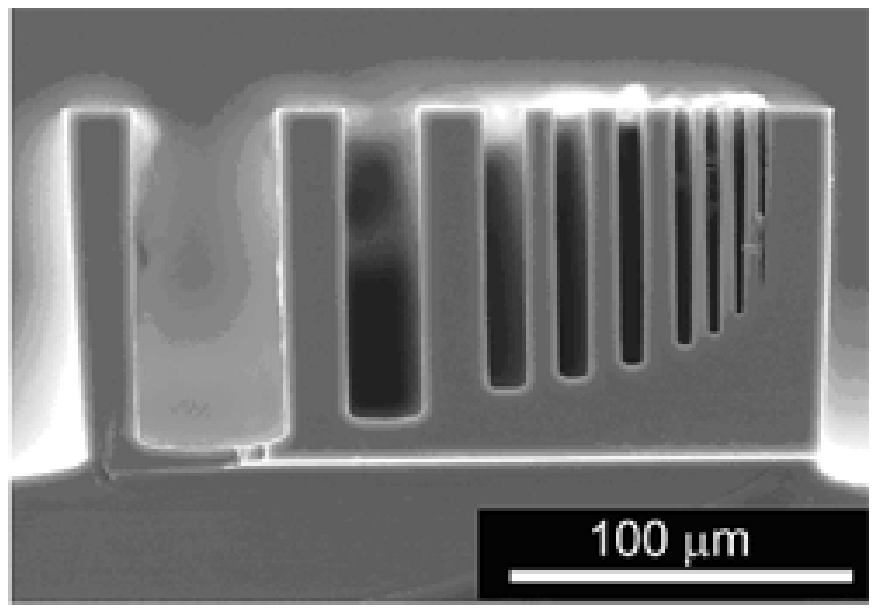}}
\label{fig12}
\end{figure}

\centerline{Fig.\:6. Cross sectional view of X-ray focusing lens [2]}

\par
\pagebreak

\begin{table}[htbp]
\begin{tabular}
{p{185pt}p{185pt}}
\centerline{\includegraphics[scale=0.75]{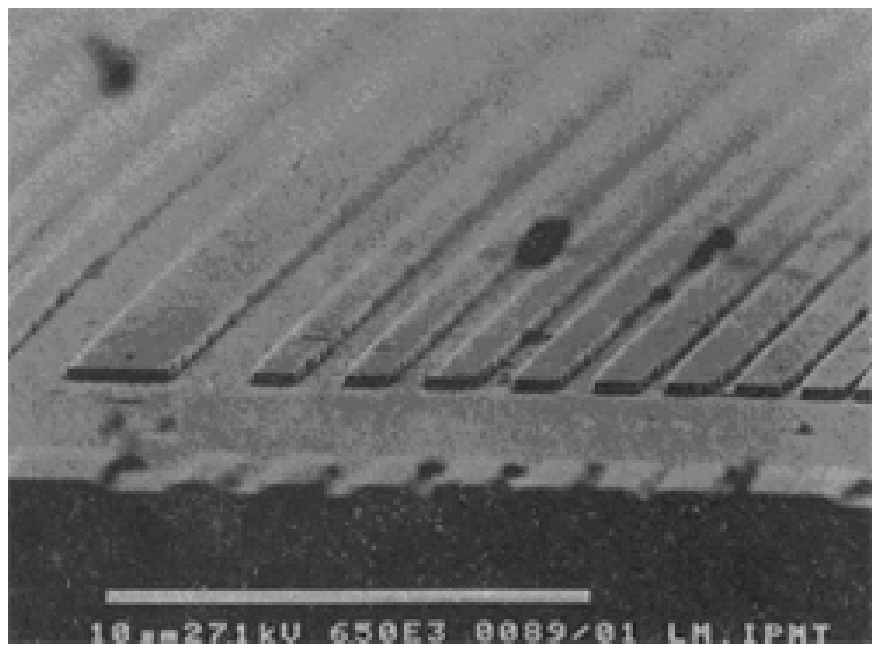}} \par &
\centerline{\includegraphics[scale=0.8]{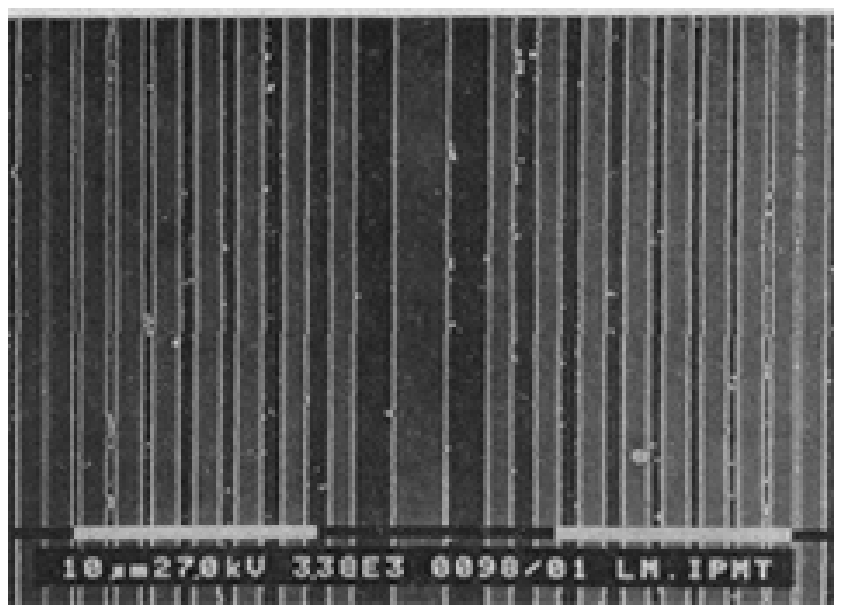}} \par  \\
\centerline{Fig. 7. Zone plate, tilted}&
\centerline{Fig. 8. Zone plate, top view} \\
\end{tabular}
\label{tab5}
\end{table}

$\phantom{a}$
\vspace*{-12mm}

It is significant, that depth decreasing depends rather on aspect ratio than
on absolute step dimensions. So this kind of defects has been seen on the
structures with elements both in micron range (fig.\:6) and in sub micron
range\linebreak (fig.\:7, 8).

It is necessary to control line width and a shape of the elements for
elliptical and round elements.
All these structures always have been studied at review stage, while with
expanding of such things production the need in inspection system will be
growing.

Has been developed both an algorithm for surface microrelief reconstruction,
based on theory of signal formation in a SEM [3], and method and attachment
to a SEM for microprofilometry [4].

\begin{figure}[htbp]
\centerline{\includegraphics[scale=1.0]{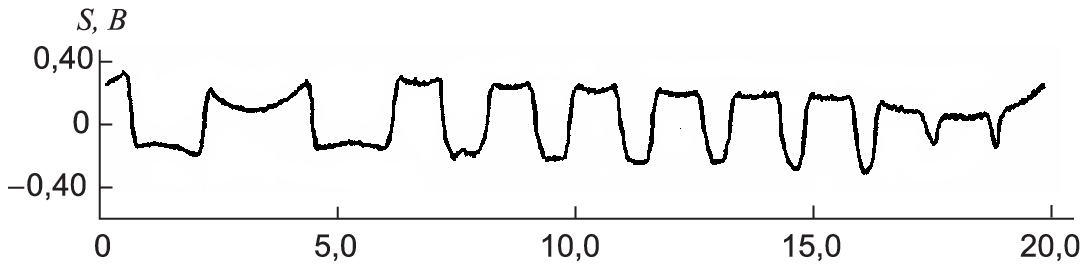}}
\label{fig12}
\end{figure}

\centerline{Fig. 9. Line scan taken from image 3; X-length in Microns}

$\phantom{a}$
\vspace*{3mm}

\begin{figure}[htbp]
\centerline{\includegraphics[scale=1.0]{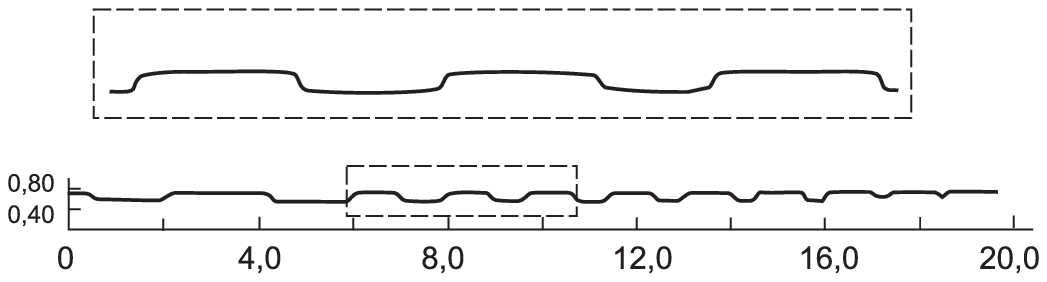}}
\label{fig12}
\end{figure}

\centerline{Fig. 10. Reconstructed surface profile along line scan}
$\phantom{a}$
\vspace*{-1mm}


It is shown in fig.\:5  a memorized signal taken from top view of structure with
"shallow" \ steps. Typically this signal is used for line width measurements.
The reconstructed profile is shown at fig.\:6. It is obvious that from this
profile can be extracted much more statistical data about structure geometry
than from usual line scan.
\vskip15pt

{\large
{\begin{center}
\textbf{II. The system requirements}
\end{center}
}}
\vskip8pt

\noindent
{\bfseries \itshape {The throughput}}
\vskip8pt

First of all we consider a correlation between beam
diameter and throughput.
As the throughput $T$ (or rather inspection speed) is understood
irradiated area for one sec.
Therefore $T=A/t$, where $A$ is usually in cm$^{2}$.

A typical value for light optics systems is $T=0,1$, which
means that wafer
200~mm in dia. Is inspected in 52~minutes.

To understand better the limitations, caused just by method of surface
irradiation (e.\,g. progressive scanning),is assumed for the beginning that we
have an "ideal" \ $e$-beam which can be focused in any small spot with
extremely high current.

For arrayed microcolumns an inspection speed can be written as
$T =(\Delta ^{2}/t ) \cdot \tN$, where $\Delta $ is a distance between
neighbour irradiated pixels which is assumed to be equal beam diameter and
resolution, $\tN$~--- quantity of microcolumns.

To estimate $N$ we assume for the beginning that  each microcolumn
needs a space $5\times 5$~mm$^{2}$ at the surface, so array for
inspection 200~mm wafer could contain 1000 microcolumns.

Dwell time $t = 100$~nsec.

The results of throughput calculations for conventional SEM ($N= 1$)
and array with 1000 microcolumns and different resolution are shown
in tab.\:1.
\vskip6pt

\noindent
{\small{Table 1.}}
\vspace*{-2mm}

\begin{table}[htbp]
\begin{tabular}
{|p{75pt}|p{54pt}|p{60pt}|p{70pt}|p{70pt}|}
\hline
\raisebox{-1.50ex}[0cm][0cm]{Resolution, nm}&
\multicolumn{2}{|p{114pt}|}{Inspection Speed,\par cm$^{2}$/sec} &
\multicolumn{2}{|p{140pt}|}{Time for wafer inspection, hr}  \\
\cline{2-5}
 &
SEM&
Array&
SEM&
Array \\
\hline 2& 0,0000004& 0,0004& 218055,5&
218,0 \\
\hline 10& 0,00001& 0,01& 8722,2&
8,7 \\
\hline 30& 0,00009& 0,09& 969,1&
0,96 \\
\hline 50& 0,00025& 0,25& 348,8&
0,35 \\
\hline 100& 0,001& 1& 87,2&
0,087 \\
\hline
\end{tabular}
\label{tab7}
\end{table}

From this simple example is possible to come to a few
important conclusions.

The main limitation on the way to combine a good resolution with high
throughput is the inspection method (progressive scanning) itself. By
increasing resolution we are decreasing quadratically the throughput, which
makes the use of SEM based system below 100~nm quite impractical. As for
arrayed microcolumns, they are potentially competitive in throughput with
light optics systems starting from resolution 30~nm.

Our consideration does not include into consideration any hint for
inspection time shortening as partial inspection and statistical methods,
when some small part of the wafer is inspected and the results further are
expanded on the whole wafer.

\vskip15pt

\noindent
{\bfseries \itshape {Signal-to-noise ratio}} {\bf(SNR)}
\vskip8pt

Differently from light optics both $e$-beam generation and emission of
secondary electrons are statistical processes.

Here we shall calculate the beam  current that must be used to give
an acceptable SNR in the recorded image or required accuracy for
quantitative measurements. We calculate SNR in the "noise
bottleneck" where numbers of signal quanta have the smallest value.
For SEM in the secondary electron mode the "noise bottleneck" is
between the sample and the collector. According
to the theory [5], if there are $N$ (on average) signal
quanta at that point, then
this will be associated with a random fluctuation $\sqrt N $. Then
$$
\mbox{SNR} =  N / \sqrt N =\sqrt N .
$$

For further calculations SNR is denoted as $n$.

$$
S = I_{b}\delta \alpha t \eqno(1),
$$
where
$I_{b}$ is primary beam current (A),  $\delta $~---
secondary emission coefficient,  $\alpha $~--- efficiency of
electrons collection by detector and $t$~--- dwell time
(sec), e.\,g. time of one
pixel irradiation.
So signal is expressed in units of charge~---
Coulombs (or A$\cdot$sec).

As $1\mbox{ C} = 6,25 \cdot10^{18 }$~electrons, the (1)
can be expressed in terms of
quantity of collected electrons:
$$
N = 6,25\cdot 10 ^{18}\cdot I_{b}\delta \alpha
t. \eqno(2)
$$

Secondary emission coefficient for Si is equal $\delta  = 0,2$ in
wide range of beam energies, and collection efficiency can vary from
0,01 to 0,9. For the most often used Everhart--Thornley detector
$\alpha$ is approximately 0,5, so for further estimations is assumed
that the product $\delta \alpha $, which shows an efficiency of the
primary electrons transformation to useful signal at detector input,
typically is equal 0,1. The dwell time is assumed to be as before:
$
t = 10^{ - 7}~sec.
$

Thus, the equation (2) can be written as:
$$
N = 6,25\cdot 10^{10}  I_{b}.  \eqno(3)
$$
Formula (3) allows one to calculate a required
primary beam current for any
in advance given SNR.

Usually is considered necessary to have $n = 3$ for detection of existence a
change of some parameter value. This smallest detectable change is often
called threshold of sensitivity or "resolution", for instance, voltage
resolution, height resolution and that sort of thing.

For good quality image that worth to be sent to a magazine $n = 16
\div  20$. In practice, when working close to the resolution limit,
a considerably higher level of noise can be tolerated in the image.
For example, if primary beam current is 1~nA, then from (3) SNR is
$$
n = 62,5^{0,5}  = 7,9,
$$
which seems to be acceptable for many practical inspection purposes.

When measuring the signal value, the accuracy
is reversely proportional to n
and error can be written as
$$
\mbox{Measurement error} = (1/n) \cdot 100\,{\%}.
$$

It is very important to remember that above estimated values of SNR related
to the background only in order to demonstrate that this background is a not
just constant value, but in part a noisy component of the informative
signal. If useful signal is less than fluctuations in a background, it will
be not seen.

It should be noted here that SNR can be deteriorated by other noisy
components which are not related to the specimen, as detector is not
positional sensitive. It accounts just an integral number of electrons at
its input. So if some of electrons are reflecting inside the SEM chamber or
inside the column and finally are coming on detector, they are considering
as useful information. That can cause confusing artifacts, so great care
should be taken of SEM construction.

\vskip15pt

\noindent
{\bfseries \itshape {Radiation damage and contamination}}
\vskip8pt

In contrast to the light quanta, the electrons interact with a
specimen. A diameter of scattering zone inside the specimen for Si
substrate can be written as $D = 0,032 E^{5 / 3}$, where $D$ is
expressed in microns and beam energy $E$ in keV. For 1~keV energy $D
= 32$~nm.

Electron scattering should be taken into consideration when
considering an informative properties of the secondary emission
signals [3,5].

However, damaging influence of electron irradiation on the solid state is
examined absolutely insufficiently, especially in the range of SEM energy
range 1 to 30 keV. Nevertheless, there are some data [6] which show that
irreversible changes in transistor structures properties (leakage current
and cutoff voltage) are determined by irradiation dose:
$$
D = I_{b}t/S, \eqno(4)
$$
where $S$ is irradiated area.
Also have been discovered changes in geometry
of masks for X-ray lithography
after irradiation with dose $D = 10^{ - 18}$~C/nm$^{2}$, what
means in terms of electrons 6,25~el/nm$^{2}$.

The above dose value seems to be small enough, however, it is equal to
sensitivity of well known PMMA electron resist. So if is inspecting a wafer
with some exposured and developed pattern on it, then after next step of
development this pattern will disappear.

Such multiple exposing and development procedure is widely used in e-beam
lithography for stitching of fragments to one big pattern. And an inspection
of the previous fragment also serves as alignment procedure for correct
positioning the following fragment and so forth.

The above examples show that a possibility of radiation damage should be
taken into consideration when developing the concept of inspection system
exactly for such sensitive things.

In the following estimations as $D$ is understood
its current value and as
$D_{m}$ is taken the most acceptable value.

It has to be understood that high resolution is unavoidably connected with
danger of sample radiation damage. As for example estimate a radiation dose
when recording a photo to confirm the SEM resolution. Typical time for image
recording is 80~sec. and picture format is $1024\times 1024$~pixels. If claimed SEM
resolution is 2~nm then pixel size is the same value and irradiated field is
about 4~square microns. If we assume that primary beam current is just 10~pA
then dose is
$$
D = 2\cdot 10^{4} \ \mu\mbox{C/cm}^{2},
$$
 which is 200 times more
than chosen $D_{m}$.

Additionally, after such sample irradiation one can see at lower
magnification that exposed area becomes dark. That happens because of
contamination caused by cracking of the long residual oil molecules into
shorter ones and adsorption of those on the surface. This adsorption layer
can hardly be removed. For example, we used that layer as a mask for ion
etching processing, and made sub micron structures with high (up to 10)
aspect ratio.

\vskip15pt

\noindent
{\bfseries \itshape {Definition of the signal contrast}}
\vskip8pt

Assume that the signal $S$ described by equation (1) is changed because of
change secondary emission coefficient $\delta $ by value $\Delta \delta $.
The change of the signal can be written as
$$
\Delta  S = I_{b}\Delta \delta \alpha  t. \eqno(5)
$$

And the contrast of the signal is defined as

$$
C = \Delta S/S. \eqno(6)
$$

It is essential to note that definition of the contrast (4) differs
from that often using in the literature $C = \Delta S/S_{\max}$.
However, the definition (4) is more convenient for analytical
applications, when by contrast is calculating height of steps or
depth of trenches on the wafer. Then the signal $S$ from flat surface
is used as a reference, which can be measured very precisely.

The reasons of local changes of secondary emission coefficient are numerous
and usually they give a name to type of contrast: topography contrast,
material (Z) contrast etc.

\vskip15pt

\noindent
{\bfseries \itshape {The correlation between
resolution, signal contrast, dose and}} \,{\bf {SNR}}
\vskip8pt

It is assumed as before that electron beam has diameter $d$ with
current $I_{b}$, and registration (and irradiation) time is $t$.

Further, it is assumed that change of signal $\Delta S$ is small
enough that noise is still defining by formula (2) as $\sqrt N$.
Therefore, SNR value n can be written as
$$
n = \Delta  S/(S)^{0,5}
$$
or, if substitute here C from (6) it can be
rewritten as
$$
n = C \cdot (S)^{0,5},  \eqno(7)
$$
or in full form
$$
n =
C\cdot (6,25 \cdot 10^{18}\cdot I_{b}\delta \alpha t)^{0,5}.
\eqno(8)
$$

Then define dose as
$D = 6,25\cdot 10^{18}\cdot I_{b} t/\left(\pi d^{2}/4\right)$ and
substitute the product $6,25\cdot 10^{18}\cdot I_{b} t$ into (8), we shall have a
final equation:
$$
d_{\min} = n/C_{\min}\cdot E\cdot(D)^{0,5}, \eqno(9)
$$
\noindent
where $E = (\delta \alpha )^{0,5}$, $d$ is expressed in nm,
and $D$~--- in
el/nm$^{2}$.

The meaning of equation (9) should be  explained in more detail,
because such an approach is used just author. At least author did
not see it somewhere else in the literature.

This formula reflects the existing  situation with transformation of
information in a SEM. It shows that each square nanometer of the
circle with diameter $d$ when irradiated with limited amount of
electrons is forming a signal (image) strongly determined quality,
that for given in advance value of SNR is able to detect some
smallest level of contrast.

\vskip15pt

\noindent
{\bfseries \itshape {The method resolution}}
\vskip8pt

As $d$ according to (9) does not depends on $e$-beam parameters, this value means not an
instrument resolution but a resolution of the method itself. To distinguish
between those two value we shall designate the method resolution as $\Delta$.
Let us explain this by example.

Assume that tolerable SNR is $n = 8$, dose $D= 6,25$~el/nm$^{2}$,
$E=0,3$ and some particle creates the contrast $C= 0,4$.
Substituting this data into (9) we have $\Delta = 8,96/ 0,3\cdot
0,4\cdot 2,5 = 29,9$~nm.
If $C=0,2$, then $\Delta =59,8$~nm. If we still are using the beam
with 29,9~ nm diameter we achieve an image with SNR $n=4$.

Now we shall try to implement achieved formula for  system
optimization.

First, is possible to optimize the throughput in some range.

Assume that beam has diameter 30 nm and dose still is $D= I_bt/d^{2}
= 6,25$~el/nm$^{2}$, the product $I_bt$ is equal to 5625 electrons,
so we can choose highest possible beam current (for its selected
diameter), which gives us the shortest dwell time and, hence,
highest throughput.

In other hand, dwell time is limited by sampling rate of analog to digital
converter (ADC) and also by timing performance of applied detector. While
modern ADCs have sampling rate over 100~MHz, the majority of detectors can
not react with such high speed.

The fastest SE detectors which are using scintillators and PMT (called often
Everhart--Thornley detector~--- ETD) unfortunately are not applicable here
because of their size. So, leaving for a while the problem of proper
detector creation, one can choose the dwell time $t = 50 \div 100$~nsec.

Then primary beam current is in the range $I = 18\div  9$~nA.

The next possibility for optimization is improving of collection efficiency
$E = (\delta \alpha )^{0,5}$.

The first coefficient $\delta $ can be increased by using beam energy close
to so-called "second crossover" \ --- the point where secondary emission yield
is equal to unity. For Si wafer this point is at the energy 2$\div $3 keV.
Detector efficiency $\alpha $ improvement is also well known and consists
mainly in increasing of collection angle $\Omega $. Thus, for the best
system value $E$ is very close to unity. In comparison with above example this
improves the system performance in 3~times.

An at the end, we'll point the way of improvement that rarely is in mind of
$e$-beam system designers. This way consists in searching an experimental
conditions which give an increasing the contrast $C$ value. We have found
first experimentally and then confirmed theoretically that contrast of the
signal from the vertical steps depends on beam energy.

From our consideration [5] follows that contrast of the signal from the step
depends on normalized height of step $h' = h/\sigma $, where $\sigma $
is a size of interaction zone between $e$-beam and specimen. For Si wafer
$\sigma \mbox{ (nm)} = 32\,E^{5 / 3}$ and $E$ expressed in keV. Here we are not
discussing all the peculiarities of surface topography reconstruction which
can be found in [4]. As for example, we assume that is measured signal from
some small Si step with height $h= 100$~nm.

Choosing the beam energy 2~keV, we calculate that $\sigma $ is about 100 nm,
and from the plot in fig the signal contrast is about 2.

Now we substitute values of contrast $C=2$ and efficiency $E$ to
formula (9) and find the resolution $\Delta _{\min} = 10/2\cdot 2,5
= 2$~nm.
\vskip16pt

\noindent
{\bfseries \itshape {Analysis of existing systems
for wafer defects inspection}}
\vskip8pt

First we estimate the idea of throughput increasing by irradiating the wafer
with large beam.

$\phantom{a}$
\vspace*{-10mm}

\begin{figure}[htbp]
\centerline{\includegraphics[scale=0.7]{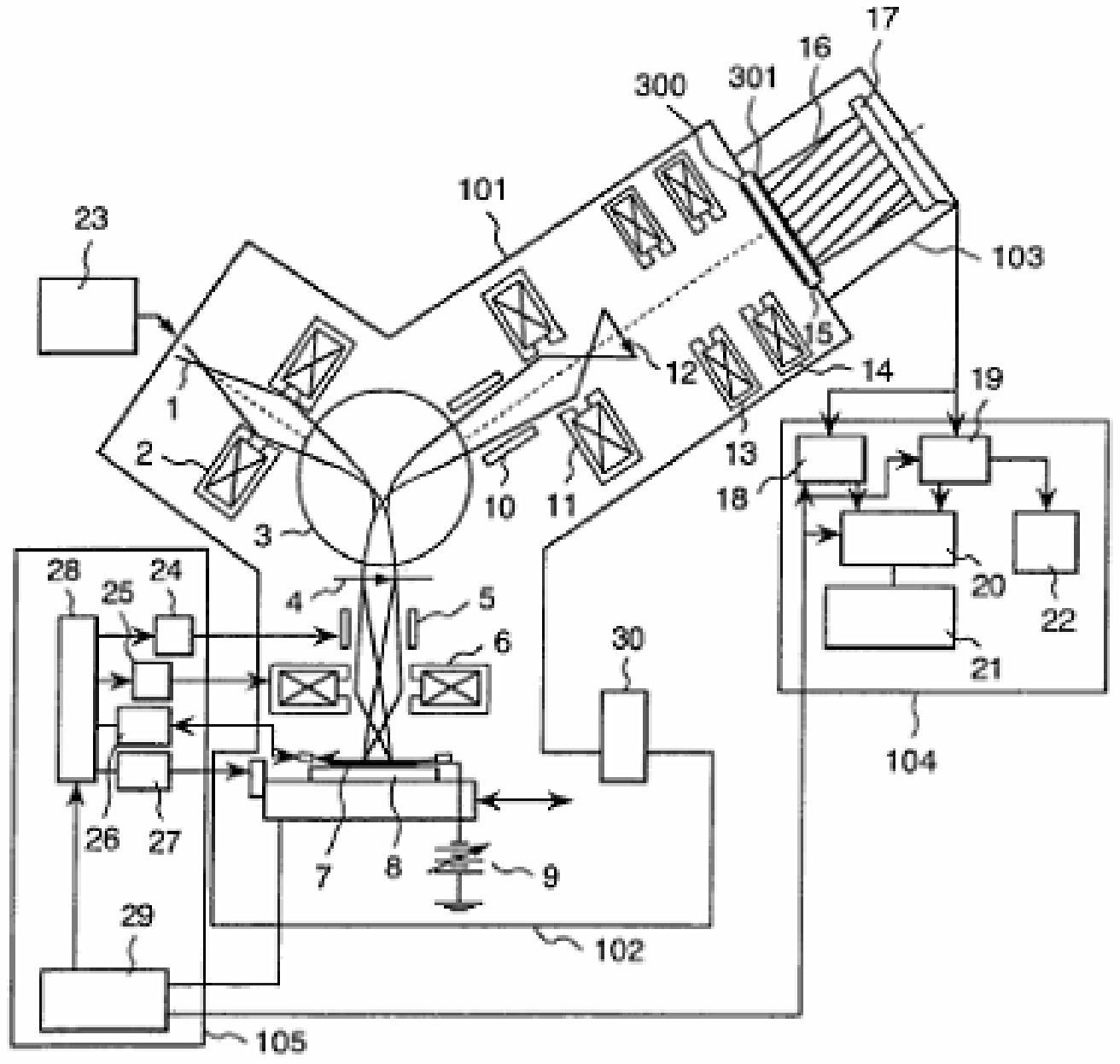}}
\label{fig16}
\end{figure}

Fig.\:11. A schematic view of the system for patterned wafer inspection [7]

$\phantom{a}$
\vspace*{-1mm}

At the fig.~11 is shown a schematic view of the system for patterned wafer
inspection according to US patent {\#}7,242, 015 B2.

The system is irradiating an area on the wafer 7 from electron source 23 via
lens 2, magnetic prism 3 and focusing lens 6. For imaging is used just a
portion of secondaries, namely reflected electrons (RE). REs are passing
through the prism 3, Wien filter 10, magnetic lens 11 and forming an
enlarged image 12 of inspected area. The image 12 can be further enlarged by
lenses 13, 14 and then registered by CCD camera 17. After image of the first
region is memorized then irradiating beam is moving to the next region by
deflector 5.

According to the claim in one shot 50~$\mu $sec is irradiated an area 100 x
100~$\mu $m with current 100~$\mu $A. Then magnification is adjusted such a
way that one pixel on CCD sensor is corresponding 0,1~$\mu $m on a specimen.
Thus, the inspection speed is 2~cm$^{2}$/sec. Signal-to-noise ratio is not
less then 10.

The resolution of the system is about 0,1~$\mu $m and inspection speed is
2~cm$^{2}$/sec. Leaving apart electron optics design, we shall
consider just an image
formation.
The authors of the invention say that image of
each pixel $0,1\times 0,1$~$\mu$m
is created by
6250 back scattered electrons (BSE).
Dose of irradiation in this case is:
$$
D= 6,25\cdot 10^{18}\cdot 100\cdot 10^{ - 6}\cdot 50\cdot 10^{ - 6}/
10^{4} = 3,125\cdot 10^{4}\mbox{ el}/ \mu \mbox{m}^{2}.
$$

\begin{figure}[htbp]
\centerline{\includegraphics[scale=0.7]{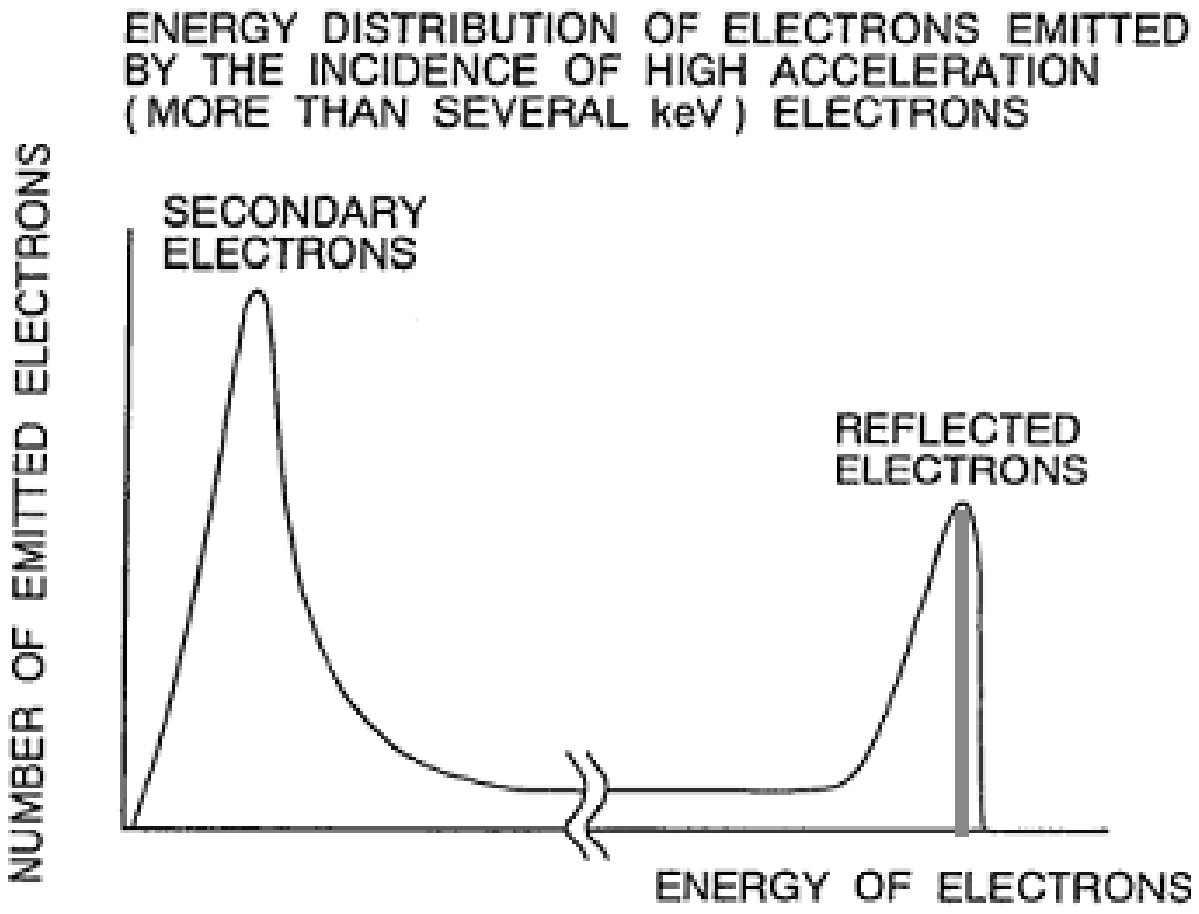}}
\label{fig17}
\end{figure}

\centerline{Fig. 12. The energy distribution of secondaries}

$\phantom{a}$
\vspace*{-1mm}

If we substitute this data into equation (9),
and assume $n=10$, $C=0,2$ and
backscattering coefficient $\eta =0,2$, then
minimal size of the square,
which is able to give image of desired
quality is: $A = 0,125$~$\mu $m, what is in a good
 agreement with authors
expectation.

However, if we would like to have a focused image it is necessary to select
from the whole BSE spectra some part with energy spread a few electron volt,
as it's shown schematically on fig.\:12 by orange strip. Otherwise the image
will be defocused because of chromatic aberration of the lens 11.

Thus, for image formation is used not $I_{b}\cdot \eta $
 number of electrons,
where $\eta =0,2$ is backscattering coefficient, but less than one per cent
of that value (fig.\:11). Energy distribution of SE and BSE.

If we assume that $n=10$, contrast is 0,2 and $E=(0,2\cdot
0,01)^{0,5}  = 1,41\times 10^{ - 2}$, we achieve that the method
resolution is:
$$
\Delta _{m}= 10/0,2\cdot 1,41\cdot 10^{ - 2}\cdot(3,125\cdot
10^{4})^{0,5} = 20 \ \mu\mbox{ m}.
$$

Thus, one can see that resolution is absolutely not satisfying. The saddest
thing here is that is not seen how it could be improved.

Now we consider one more embodiment of defect inspection system (Fig.\:13)
It is a good example of both whole single beam system design and its
optimization for highest possible throughput.

\begin{figure}[htbp]
\centerline{\includegraphics[scale=1.3]{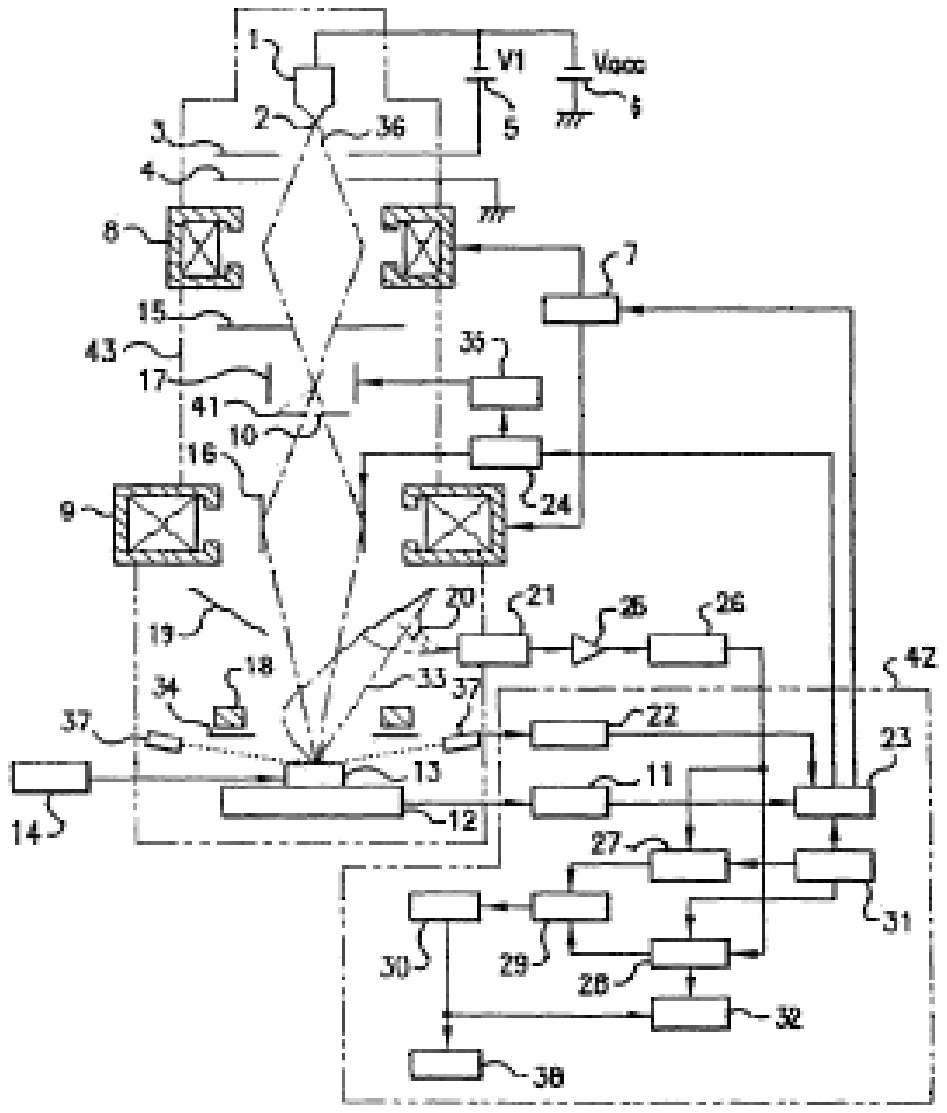}}
\label{fig18}
\end{figure}

\centerline{Fig. 13. The inspection system according to US patent 7,232,996 B2 [8]}

$\phantom{a}$
\vspace*{-1mm}

The key characteristics are:

resolution~--- 0,1~$\mu$m;

beam current~--- 100~nA;

dwell time~--- 10~nsec.

Hence, current dose is $D = 6,25$~el/$\mu $m$^{2}$ or 10 $\mu
$C/cm$^{2}$; the method resolution $\Delta   = 0,14$~$\mu $m.
However, that combination  of parameters can not be improved any
further.

Assume for example, that beam is focused into spot 0,05~$\mu $m. Then beam
current should also decrease about 25~nA, so dose remains the same.
Therefore, the method resolution also does not change and image quality is
getting worse. That happens because number of secondaries from pixel becomes
4~times smaller, hence SNR is 2~times lower. In order to achieve the same
picture quality (but with 2~times better resolution) it is necessary to
increase irradiation time since 10~nsec to 40~nsec.

Nevertheless, this system is a good example to consider all the components,
not just probe forming system.
Wafer stage 12 is continuously moving that allows to avoid "dead" \ time. The
stage is equipped with a length measuring unit 11.
The system is equipped with height measuring unit 37.

\vskip15pt

{\large
{\begin{center}
\textbf{The discussion}
\end{center}
}}
\vskip8pt

Achieved relationship between signal, SNR and method resolution enables an
estimation of the inspection system information capability at whole and its
applicability for particular work.

In frames of existing system is possible to optimize its operational
conditions (magnification, dwell time, accelerating voltage) in order to
extract as much information as possible.

Analysis of existing single beam defects inspection systems shows that their
practical implementation is limited by resolution 0,1~$\mu$m and this can
not be improved any further by electron optical means.

The systems for topographycal defect inspection in the range of 10 - 30 nm
can be realized just with array of microcolumns. Fortunately, electron
optics design in that range is relatively simple.

The only practical implementation for which can be really needed ultra high
resolution about 2 nm is line width measuring (LWM) and microrelief profile
reconstruction. However, besides an advanced electron optical design for
successeful application of LWM system at the inspection stage, an additional
means as statistical data processing. image recognition and others similar
techniques are indispensable.

\vskip15pt

{\large
{\begin{center}
\textbf{References}
\end{center}
}}
\vskip8pt

\begin{itemize}
\item[\hbox{[1]}]\textit{ Frank Y.H. Fan}. Method and an apparatus
for reviewihg voltage contrast
defects in semiconductor wafers~// United States Patent. Patent No.: US 7,132,301 B1.
\item[\hbox{[2]}]\textit{ V. Yunkin, M. Grigoriev, S. Kuznetsov,
A. Snigirev, I. Snigireva}. Planar parabolic refractive lenses for hard
x-rays:technological aspects of fabrication~// Proc. of SPIE. Vol. 5539. P. 226-234.
\item[\hbox{[3]}] \textit{ V.V. Aristov, N.N. Dremova,
A.A. Firsova, V.V. Kazmiruk, A.V. Samsonovich,
N.G. Ushakov, S.I. Zaitsev}. Signal formation of backscattered electrons  by microinhomogeneities and surface relief in a SEM // Scanning. Vol. 13. (1990) P. 15-22.

\item[\hbox{[4]}]\textit{ V.V. Kazmiruk, V.I. Mjasnikov,
T.N. Savitskaja, I.S. Stepanov}. Attachment to SEM for
local profilometry of surface~// Izv. Akademii nauk SSSR. Ser. Fiz. V. 56. № 3. (1992)  P. 58-63.

\item[\hbox{[5]}]\textit{ V.V. Kazmiruk, T.N. Savitskaja,
I.S. Stepanov, A.A. Firsova}. Research of forming in SEM  the backscattered electrons  signal by surface microrelief // Изв Izv. Akademii nauk SSSR. Ser. Fiz. V. 54. № 2. (1990) P. 227-231.

\item[\hbox{[6]}] \textit{V.V. Kazmiruk, V.I. Mjasnikov,
T.N. Savitskaja, I.S. Stepanov}. Measuring potencial relief
on the structures with submicron size elements~// Izv. Akademii
nauk SSSR. Ser. Fiz. V. 54. № 2. (1990) P. 305.

\item[\hbox{[7]}]\textit{Shinada et al}. Patterned wafer
inspection method and apparatus therefore~// United States Patents.
Patent No.: US 7,242,015 B2.

\item[\hbox{[8]}]\textit{Iwabuchi et al}. Method and an
apparatus of an inspection system using an electron beam~// United States
Patent. Patent No.: US 7,232,996 B2.
\end{itemize}

\end{document}